\documentclass[aps,pre,twocolumn,superscriptaddress,showpacs]{revtex4-1}
\usepackage{float}
\usepackage{times}
\usepackage[dvips]{color}
\usepackage{mathtools}
\usepackage{dcolumn}
\usepackage{bm}
\usepackage{amsmath}
\usepackage{amssymb}
\usepackage{hyperref}

\renewcommand{\d}{\mathrm{d}}
\newcommand{\ds}{D_{\mathrm{s}}}

\newcommand{\bs}{\beta_{\mathrm{s}}}

\begin{document}

\title{Small-angle scattering from the Cantor surface fractal on the plane and the Koch snowflake}

\author{A. Yu. Cherny}
\email[e-mail:~]{cherny@theor.jinr.ru}
\affiliation{Joint Institute for Nuclear Research, Dubna 141980, Russian Federation}
\affiliation{Center for Theoretical Physics of Complex Systems, Institute for Basic Science (IBS), Daejeon 34051, Republic of Korea}

\author{E. M. Anitas}
\affiliation{Joint Institute for Nuclear Research, Dubna 141980, Russian Federation}
\affiliation{Horia Hulubei National Institute of Physics and Nuclear Engineering, RO-077125 Bucharest-Magurele, Romania}

\author{V. A. Osipov}
\affiliation{Joint Institute for Nuclear Research, Dubna 141980, Russian Federation}

\author{A. I. Kuklin}
\affiliation{Joint Institute for Nuclear Research, Dubna 141980, Russian Federation}
\affiliation{Laboratory for Advanced Studies of Membrane Proteins, Moscow Institute of Physics and
Technology, Dolgoprudniy, Russian Federation}

\date{\today}

\begin{abstract}
The small-angle scattering (SAS) from the Cantor surface fractal on the plane and Koch snowflake is considered. We develop the construction algorithm for the Koch snowflake, which makes possible the recurrence relation for the scattering amplitude. The surface fractals can be decomposed into a sum of surface mass fractals for arbitrary fractal iteration, which enables various approximations for the scattering intensity. It is shown that for the Cantor fractal, one can neglect with a good accuracy the correlations between the mass fractal amplitudes, while for the Koch snowflake, these correlations are important. It is shown that nevertheless, the correlations can be build in the mass fractal amplitudes, which explains the decay of the scattering intensity $I(q)\sim q^{D_{\mathrm{s}}-4}$ with $1 < D_{\mathrm{s}} < 2$ being the fractal dimension of the perimeter. The curve $I(q)q^{4-D_{\mathrm{s}}}$ is found to be log-periodic in the fractal region with the period equal to the scaling factor of the fractal. The log-periodicity arises from the self-similarity of \textit{sizes} of basic structural units rather than from correlations between their distances. A recurrence relation is obtained for the radius of gyration of Koch snowflake, which is solved in the limit of infinite iterations. The present analysis allows us to obtain additional information from SAS data, such as the edges of the fractal regions, the fractal iteration number and the scaling factor.
\end{abstract}


\maketitle

\section{\label{sec:Introduction}Introduction}
The small-angle scattering (SAS) of waves (e.g. neutrons, X-rays, light)~\cite{guinier55:book,svergun87:book,brumberger95:book,lindner02:book} is an important non-destructive method in determining the structural properties (internal structure, shape, size, positional correlations, average spatial arrangement, molecular weight, fractal dimension) of fractal and/or disordered systems (polymers, complex fluids, aggregates, colloids) at nano and microscales~\cite{martin87,schmidt91,beaucage96,mortensen96,fritz06,erko12}. In particular, by using the framework provided by deterministic (exact self-similar) fractals, it has been shown more recently that the range of structural properties which can be extracted can be significantly extended to include additional information, such as the scaling factor, iteration number or the number of particles constituting the fractal~\cite{chernyJACR10,chernyPRE11}. These information are usually extracted from a double logarithmic plot of the normalized elastic cross section per unit volume of the sample (scattering intensity) $I(q) \equiv (1/V^{'}) d \sigma/ d \Omega$ plotted versus the scattering wave vector $q=(4\pi /\lambda)\sin \theta$  ($\theta$ is half the scattering angle and $\lambda$ is the wavelength of the incident radiation) which describes, through a Fourier transform, the spatial density-density correlations of the system. Therefore, the information obtained by using SAS coupled with the theoretical framework provided by the fractal geometry~\cite{mandelbrot83:book,gouyet96:book} allows us to have a better understanding of the structural properties of such systems.

An important characteristic, which makes SAS a unique tool in analyzing experimental data from fractal systems is the possibility to differentiate between ``mass" and ``surface" fractals~\cite{bale84,teixeira88}. The difference arise from the value of the scattering exponent $\tau$ of the simple power-law SAS intensity:
\begin{equation}
I(q)\propto q^{-\tau},
\label{eq:sas}
\end{equation}
where the scattering exponent can be written in the following way
\begin{equation}
\tau = \begin{dcases}
   D_{\mathrm{m}}, &\mathrm{for~mass~fractals},\\
   2d-D_{\mathrm{s}}, &\mathrm{for~surface~fractals}.
   \end{dcases}
  \label{eq:tau}
\end{equation}
Here, $d$ is the topological dimension of the space into which the fractal is embedded, $D_{\mathrm{m}}$ is the mass fractal dimension satisfying the condition $0<D_{\mathrm{m}}<d$, and $D_{\mathrm{s}}$ is the surface fractal dimension, satisfying $d-1<D_{\mathrm{s}}<d$. For three-dimensional space ($d=3$), this leads to a simple interpretation of SAS experimental data: if the power-law exponent $\tau <3$, the measured  sample is a mass fractal, while if $3<\tau<4$ then the sample is a surface fractal. One can adopt a simple descriptive definition of the Hausdorff dimension $D$ of a set as the exponent in the relation $N\varpropto (1/a)^D$ for $a\to 0$, where $N$ is the minimum number of open sets of diameter $a$ needed to cover the set. For a ``usual"  globular object like ball, the Hausdorff dimensions of volume and surface are equal to 3 and 2, respectively.

Sometimes a succession of simple power-law decays with different exponents can be observed in SAS data, which can be explained by the presence of a fractal structures at different scales in monophase~\cite{beaucage96,anitasEPJB14} and multiphase~\cite{chernyJACR14} systems.

In the previous publications~\cite{cherny110,chernyJACR10,chernyPRE11}, the fractal models with controllable dimension are suggested, for which the scattering amplitude is known analytically. Being exactly solvable, the models are quite convenient and effective to study and check many statements and estimations accepted a long time ago from general considerations.

More recently, it was suggested that \emph{any surface fractal is composed of mass fractals of the same fractal dimension}, and thus, the scattering amplitude of the surface fractal is the sum of the amplitudes of composing mass fractals~\cite{chernyARXIVE16}. This has been verified for the $3D$ Cantor-like surface fractal, which is build from Cantor dusts at various iterations. Some caveats  should be used here. First, alternatively, the fractal support of a surface fractal can be constructed by a \textit{subtraction} of mass-fractal iterations, because adding measure to a fractal support is equivalent to subtracting the same measure from the complementary set, and in accordance with Babinet's principle, two complementary sets give the same diffraction pattern. Second, mathematically, the limit of infinite number of iterations of composing mass fractal might not exist. However, in a realizable surface fractal sample, the building mass fractals always have finite iterations with well-pronounced scaling properties, which alone are important in the SAS from the surface fractal sample.

It has been shown~\cite{chernyARXIVE16} that the ``rough structure" of the scattering intensity can be explained in terms of power-law distribution of sizes of objects composing the fractal. The power-law decay $I(q) \propto q^{D_{\mathrm{s}}-6}$ is realized as a non-coherent sum of scattering amplitudes of $3D$ objects composing the fractal and obeying a power-law distribution $dN(r) \propto r^{\tau}dr$, with $D_{\mathrm{s}} = \tau-1$. We mean by rough structure that not all minima and maxima superimposed on the power-law decay appear within this approximation.

Here, we apply the above findings to the well-known $2D$ Koch snowflake and Cantor surface fractals. It is shown that for the Cantor surface fractal, the approximation of independent composite units works fairly well, while for the Koch snowflake it works only ``roughly", that is, it does not reproduce the fine structure of minima and maxima of the scattering intensity. However, it is possible to develop the other approximations with the help of the decomposition of surface fractal into mass fractals~\cite{chernyARXIVE16}. The Koch snowflake gives us an example where the difference is quite pronounced between the sum of intensities of composing units and the sum of intensities of pairs of consecutive amplitudes (see Sec.~\ref{sec:koch} below).

On the other hand, the suggested methods can be interesting also from the mathematical point of view. We develop an algorithm of construction of the Koch snowflake, which enables us to obtain the exact and quite aesthetic recurrence relation for the scattering amplitude (the latter is nothing but the Fourier transform of its support) and by means of this, to obtain the recurrence relation for the radius of gyration. The latter quantity is obtained explicitly for the ideal fractal, that is, in the limit of infinite number of iterations.

This paper is structured as follows. The general remarks about the SAS scattering from fractals is given in Sec.~\ref{sec:Theory}. In the sections \ref{sec:dfs} and \ref{sec:koch}, the SAS intensities are calculated in momentum space and analyzed. In Sec.~\ref{sec:koch}, an algorithm for constructing the Koch snowflake is considered, which is suitable to obtain the recurrence relation for the scattering amplitude. It is shown that the curve $I(q)q^{4-D_{\mathrm{s}}}$ with $1 < D_{\mathrm{s}} < 2$, is log-periodic with the period equal to the scaling factor of the fractal. In this section, the recurrence relation is obtained for the radius of gyration of Koch snowflake. In the Conclusion, the main results and prospects are discussed.

\section{\label{sec:Theory}Theoretical background}

We consider single scattering from a large number of arbitrarily oriented surface fractals whose positions are uncorrelated. In the two-phase system that consists of fractals of volume $V$ and concentration $n$ and a surrounding medium, the differential cross section per unit volume of the sample (scattering intensity) is given by~\cite{svergun87:book}
\begin{equation}
I(q) \equiv \frac{1}{V^{'}}\frac{\mathrm{d}\sigma}{\mathrm{d}\Omega} = n \left| \Delta \rho \right|^{2} V^{2} \left\langle \left| F(\bm{q}) \right|^{2} \right\rangle.
\label{eq:dcc}
\end{equation}
Here $\Delta\rho$ is the scattering contrast between the fractal support and surrounding madium, $V'$ is the total volume irradiated,
\begin{equation}
F(\bm{q})=\int_{V}e^{-i\bm{q} \cdot \bm{r}}\mathrm{d}\bm{r}/V,
\label{eq:normff}
\end{equation}
while the symbol $\langle \cdots \rangle$ stands for ensemble averaging over all orientations.

Let us consider a mass fractal of fractal dimension $D_{\mathrm{m}}$ with the total length $L$, which is composed of $p$ basic units. Each init is of size $l$ and separated by distances $d$, with $l \lesssim d \lesssim L$. Since the number of basic units is of the order of $(L/d)^{D_{\mathrm{m}}}$, the normalized form factor is given by
	\begin{equation}
	\left\langle \left| F_{\mathrm{m}}(\bm{q}) \right|^{2} \right\rangle \simeq
	\begin{dcases}
	1, &q \lesssim 2\pi/L,\\
	(qL/2\pi)^{-D_{\mathrm{m}}}, &2\pi/L \lesssim q \lesssim 2\pi/d,\\
	(L/d)^{-D_{\mathrm{m}}}, &2\pi/d \lesssim q \lesssim 2\pi/l,\\
	(L/d)^{-D_{\mathrm{m}}}(ql/2\pi)^{-4}, &q \gtrsim 2\pi/l,
	\end{dcases}
	\label{eq:ffmassf}
	\end{equation}
which clearly shows the presence of four main regions: Guinier, fractal, plateau and Porod.

For the case of a surface fractal (of fractal dimension $D_{\mathrm{s}}$) composed of basic units of largest size $L_{0}$, and smallest size $l$, the normalized form factor becomes
	\begin{equation}
	\left\langle \left| F_{\mathrm{s}}(\bm{q}) \right|^{2} \right\rangle \simeq
	 \begin{dcases}
	   1, &q \lesssim 2\pi/L_0,\\
	   (qL_0/2\pi)^{D_{\mathrm{s}}-6}, &2\pi/L \lesssim q \lesssim 2\pi/l,\\
	   (L_0/l)^{D_{\mathrm{s}}-6}(ql/2\pi)^{-4}, &q \gtrsim 2\pi/l.
	 \end{dcases}
	\label{eq:ffsurfacef}
	\end{equation}

In the above expressions, we make some simplifications, for instance, the Guinier region is not parabolic as it should be. We are most interested in the fractal regions, for which our approximations work (see below). While for a mass fractal, the fractal region arises due to spatial correlations of the basic units~\cite{chernyPRE11}, for surface fractals it arises due to the power-law polydispersity in their sizes~\cite{chernyARXIVE16}. As a consequence, the fractal region of mass fractals is given by the maximal and minimal distances between centers of basic units, while for surface fractals, is given by the largest and smallest dimensions of the basic units. Finally, the plateau between $2\pi/d \lesssim q \lesssim 2\pi/l$ in Eq.~\eqref{eq:ffmassf} can be considered also as a Guinier region for the basic units, since spatial correlations between different units does not play an important role.

At $n$-th iteration, the normalized scattering amplitude for a mass fractal is
\begin{equation}
F_{n}^{(\mathrm{m})}(\bm{q}) = F_{0}(\beta_{\mathrm{s}}^{n}qr_{0})G_{1}(\bm{q})G_{1}(\beta_{\mathrm{s}}\bm{q}) \cdots G_{1}(\beta_{\mathrm{s}}^{n-1}\bm{q}),
\label{eq:normamp}
\end{equation}
where $F_{0}(q)$ is the form factor of the basic fractal unit, $G_{1}(\bm{q})$ is the generative function depending on the relative positions of the fractal units inside the fractal, and $\beta_{\mathrm{s}}$ is the scaling factor of the fractal.

Thus, since the surface fractal is the sum of mass Cantor fractals at various iterations, we shall add the amplitudes of the mass fractal iterations, and normalize the result to one at $q = 0$. Then, for a $3D$ Cantor-like surface fractal we have~\cite{chernyARXIVE16,chernyRJP15}
\begin{equation}
F_{m}^{(\mathrm{s})}(\bm{q}) = \frac{1-k\beta_{\mathrm{s}}^3}{1-(k\beta_{\mathrm{s}}^3)^{(m+1)}}\sum_{n=0}^{m}(k\beta_{\mathrm{s}}^3)^nF_{n}^{(\mathrm{m})}(\bm{q}),
\label{eq:ampsf}
\end{equation}
where $k$ is the number of balls of radius $r_{1}=\beta_{\mathrm{s}}r_{0}$ which replace the ball of radius $r_{0}$ at zero-th iteration (initiator). Then, the final expression for the scattering intensity is given by (see Eq.~\ref{eq:dcc})
\begin{equation}
I_{m}^{(\mathrm{s})}(q) = I_{m}^{(\mathrm{s})}(0)\left\langle \left| F_{m}^{(\mathrm{s})}(\bm{q}) \right|^{2} \right\rangle,
\label{eq:finalint}
\end{equation}
with $I_{m}^{(\mathrm{s})}(0)=n\left| \Delta \rho \right|^{2}V_{m}^{2}$, where $V_{m}$ is the total volume of surface fractal at $m$-th iteration.

In two dimensional space, the scattering intensity $I(q)$ of a set is calculated by means of averaging the squared scattering amplitude $S^2 \langle|F(\bm{q})|^2\rangle$ with respect to the polar angle in the plane, where $S$ is the area of the set. The normalized amplitude (\ref{eq:normff}) is calculated by integration in the plane: $F(\bm{q})=\int_S e^{-i\bm{q} \cdot \bm{r}}\mathrm{d}\bm{r}/S$. The intensity decays for large scattering vectors as $I(q) \propto q^{4-D_{\mathrm{s}}}$ with $1 < D_{\mathrm{s}}<2$, see Eq.~(\ref{eq:tau}). Here the ``surface" dimension $D_{\mathrm{s}}$ is nothing else but the Hausdorff dimension of perimeter bounding the fractal support in the plane. For this reason, it would be more natural to talk about a ``perimeter" fractal in the plane, but we use, nevertheless, the well-known ``three-dimensional" terminology~\cite{chernyARXIVE16} and adopt the notation $D_{\mathrm{s}}$. The perimeter of a ``usual" object, like disk or rectangle, is a one-dimensional line ($D_{\mathrm{s}}=1$).

\section{\label{sec:dfs}Cantor-like surface fractal}

\subsection{\label{subsec:Construction}Construction and properties}

The construction process of $2D$ Cantor-like surface fractal is similar to that of $3D$ version~\cite{chernyARXIVE16} in the sense that one follows a "top-down" approach in which an initial structure is repeatedly divided (by a single scaling factor) into a set of smaller structures of the same type according to a given rule which is kept the same from one iteration to the next one.

\begin{figure}
\begin{center}
\includegraphics[width=\columnwidth]{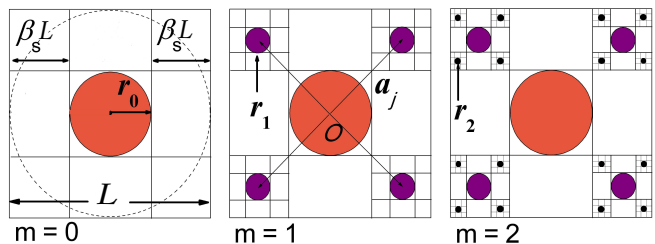}
\end{center}\caption{
(Color online) Construction of Cantor-like surface fractals at $m = 2$ as a sum of Cantor mass fractals at: $m = 0$ (disk of radius $r_0$; orange), $m = 1$ (disks of radii $r_1 = r_0 \beta_{\mathrm{s}}$; violet) and $m = 2$ (disks of radii $r_2 = r_0 \beta_{\mathrm{s}}^{2}$; black). The vectors $\textbf{a}_{j}$, $j=1,2,3,4$ connect the center of disk of radius $r_{0}$ with the centers of disks of radii $r_{1}$. The components of these vectors are $\pm {(1-\beta_\mathrm{s})}{L}/{2}$.}
\label{fig:Cantorsurfacefractal}
\end{figure}

The Cantor surface fractal is constructed as a sum of mass generalized Cantor fractals (GCF), which are suggested and discussed in detail in Refs.~\cite{chernyJACR10,chernyPRE11}. The GCF is also called Cantor dust. Let us recall the construction algorithm for the GCF. We start with a square of edge ${L}$ and choose a Cartesian system of coordinates with the origin in the square center, and the axes parallel to the cube edges. The zeroth iteration (called initiator) is a disk of radius $r_{0}$ in the origin.
The iteration rule (generator) is to replace the disk with $k$ smaller disks ($k=4$) of radius $r_{1}=\beta_{\mathrm{s}}r_{0}$, where the parameter $\beta_{\mathrm{s}}$, called scaling factor, obeys the condition $0 <\beta_{\mathrm{s}} <1/2$. The centers of the four disks of radius $r_{1}$ are shifted from the origin by the four vectors
\begin{equation}
\bm{a}_{j}=\frac{1-\beta_\mathrm{s}}{2}{L}\left\{ \pm 1, \pm 1 \right\}
\label{eq:shifts}
\end{equation}
with all the combinations of the signs. The next iterations are obtained by performing an analogous operation to each of $k$ balls of radius $r_{1}$, and so on
(see Fig.~\ref{fig:Cantorsurfacefractal}). The fractal dimension of the Cantor dust (mass Cantor fractal) is given by \cite{chernyJACR10}
\begin{equation}\label{dimmass}
D_{\mathrm{m}} =-\ln k/\ln\beta_{\mathrm{s}}
\end{equation}
with $k=4$ for the Cantor dust in two dimensions. It lies within $0<D_{\mathrm{m}}<2$. 

The $m$-th iteration of the Cantor-like surface fractal is built as a \emph{sum} of the Cantor dusts of iterations from zero to $m$, see Fig.~\ref{fig:Cantorsurfacefractal}. In order to avoid the overlapping between the different iterations of the Cantor dust, the initial radius should be restricted: $r_0\leqslant {L}(1-2\beta_{\mathrm{s}})/2$. Thus, the largest size contained in Eq.~(\ref{eq:ffsurfacef}) is equal to $L_0=2r_0$. By the construction, the initial length ${L}$ is nothing else but the size of the surface fractal if $m$ is big enough. The main difference between the Cantor mass and surface fractals is that, at a given iteration, the mass fractal consists of subunits with the same size, while the surface fractal consists of subunits with different sizes. As we shall see below, this property is responsible for the scattering behavior $I(q)\propto q^{D_{\mathrm{s}}-4}$. The difference is apparent from Fig.~\ref{fig:Cantorsurfacefractal}.

At the $m$-th iteration, the two-dimensional Cantor-like surface fractal is composed of $N_m=1+k+k^2+\cdots+k^m$ balls
\begin{equation}\label{Nm}
N_m=(k^{m+1}-1)/(k-1)
\end{equation}
(with $k=4$), whose radii and surface areas are distributed in the following way. One disk of radius $r_0$ has area $\pi r_{0}^{2}$, $k$ disks of radius $r_{1}=\beta_{\mathrm{s}}r_{0}$ have the area $k \pi r_{1}^{2}$, $k^{2}$ disks of radius $r_{2}=\beta_{\mathrm{s}}^{2}r_{0}$ have the area $k^{2} \pi r_{2}^{2}$), and so on. Then, the total area of surface fractal at $m$-th iteration is given by
\begin{equation}\label{totSSF}
S_{m}=S_0\frac{1-(k\beta_{\mathrm{s}}^2)^{m+1}}{1-k\beta_{\mathrm{s}}^2}
\end{equation}
with the volume of zero iteration $S_0=\pi r_{0}^{2}$. Because of the inequality $k\beta_{\mathrm{s}}^2<1$, the total area (\ref{totSSF}) is finite in the limit $m\to\infty$, and then the Hausdorff dimension of the fractal \emph{surface} is equal to 2.

The contribution of the initiator ($m=0$) to the Hausdorff dimension of the total \emph{perimeter} of the Cantor-like fractal is obviously equal to 1, which yields the lower limit for the surface dimension, while the contribution of the $m$-th mass iteration for $m\to \infty$ is given by the fractal dimension (\ref{dimmass}). Then we arrive at the the Hausdorff (fractal) dimension of the total \emph{perimeter} of the Cantor-like fractal
\begin{equation}
D_{\mathrm{s}} =\begin{dcases}
   1, &\text{for}\ 0<\beta_{\mathrm{s}}\leqslant 1/k,\\
   -\ln k/\ln\beta_{\mathrm{s}}, &\text{for}\ 1/k\leqslant\beta_{\mathrm{s}}<1/2.
   \end{dcases}
\label{eq:sfddeter}
\end{equation}
The threshold  value $\beta_{\mathrm{s}}=1/k$ corresponds to $D_{\mathrm{m}}=1$ in Eq.~(\ref{dimmass}), which yields $\beta_{\mathrm{s}}=0.25$ for $k=4$. When the scaling factor $\beta_{\mathrm{s}}$ smaller than this value, the total surface of the fractal is finite even in the limit $m\to\infty$. As expected \cite{bale84,martin87,schmidt91}, the surface Hausdorff dimension satisfies the condition $1 <  D_{\mathrm{s}} <2$.

\subsection{\label{subsec:monoformfactor}Fractal form factor}
In two dimensions, the number of disks in the first iteration of the mass fractal is $k=4$ and their radius is $r_{1}=\bs r_0$. The form factor of a disk of unit radius is given by \cite{guinier55:book}
\begin{equation}
F_{0}(z)=2J_{1}(z)/z,
\label{2Dff}
\end{equation}
where $J_{1}(z)$ is the Bessel function of the first kind of order one. The generative function is $G_{1}(\bm{q}) \equiv \cos(u q_{x})\cos(u q_{y})$, where $u \equiv {L}(1-\beta_{\mathrm{s}})/2$.

By neglecting the correlations between the amplitudes of different mass fractal iterations (that is  $\langle F_{n}^{\mathrm{(mf)}}(\bm{q})F_{j}^{\mathrm{(mf)}}(\bm{q})\rangle \simeq 0$ for $n\neq j$) when $q\gtrsim 2\pi/r_{nj}$ with $r_{nj}$ being a typical distance between balls in the $n$th and $j$th mass fractal iterations, and using Eq.~(\ref{eq:finalint}), the scattering takes the form~\cite{chernyARXIVE16}
\begin{align}
I_{m}^{\mathrm{(sf)}}&(q)/I_{m}^{\mathrm{(sf)}}(0)=\langle |F_{m}^{\mathrm{(sf)}}(\bm{q})|^2\rangle \nonumber\\
&\simeq \frac{(1-k\beta_{\mathrm{s}}^2)^2}{\big(1-(k\beta_{\mathrm{s}}^2)^{m+1}\big)^2}\sum_{n=0}^{m}(k\beta_{\mathrm{s}}^2)^{2n}\langle |F_{n}^{\mathrm{(mf)}}(\bm{q})|^2\rangle,
\label{eq:SFincohmass}
\end{align}

Following similar arguments as in Ref.~\cite{chernyARXIVE16} we can rewrite the incoherent sum of scattering intensities of disks as:
\begin{equation}
I^{(\mathrm{s})}(q) \simeq I_{0}(q)+\beta_{\mathrm{s}}^{4-D_{\mathrm{s}}}I_{0}(\beta_{\mathrm{s}}q)+(\beta_{\mathrm{s}}^{4-D_{\mathrm{s}}})^{2}I_{0}(\beta_{\mathrm{s}}^{2}q)+\cdots,
\label{eq:incohsas}
\end{equation}
where $I_{0}(q)$ is the scattering intensity of the central scattering disk.

\begin{figure}
\begin{center}
\includegraphics[width=\columnwidth]{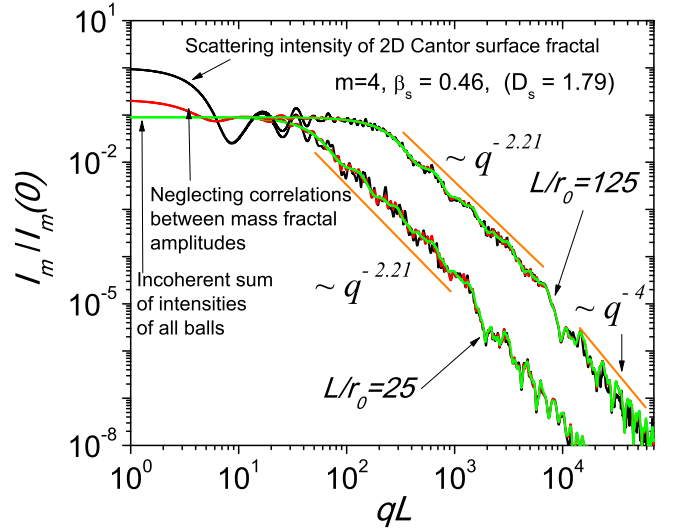}
\end{center}\caption{(Color online) Scattering intensities from two-dimensional Cantor fractals. Note that the condition ${L}/r_0 \geqslant 2/(1-2\beta_{\mathrm{s}})$ guarantees the absence of overlapping between the structural units of the Cantor-like surface fractal. One can observe a good agreement between the exact formula, the approximation neglecting the correlations between the mass fractal amplitudes, and completely incoherent sum of intensities of the disks composing the fractal.}
\label{fig:fig2}
\end{figure}

The figure \ref{fig:fig2} shows that the scattering intensity of a surface fractal is realized approximately as a \textit{non-coherent sum} of intensities of a system of disks. One can see that in the fractal region $\pi/r_0 \lesssim q \lesssim \pi/r_{m}$, we have a good coincidence between exact formula (\ref{eq:finalint}), the approximation  (\ref{eq:SFincohmass}) neglecting the correlations between mass fractal amplitudes, and completely incoherent sum of intensities of the disks (\ref{eq:incohsas}), which are discussed in detail in Ref.~\cite{chernyARXIVE16}. The latter approximation is given by Eq.~(\ref{eq:incohsas}) but with $I_0(q)\equiv n |\Delta\rho|^{2} \pi^2 r_0^4 F_0^2(q r_0)$  and the exponent $4-\bs$. The different curves correspond to different sizes of the initial disk radius $r_0$. In both cases, as expected, we can observe the presence of the four main regions of scattering intensities: Guinier, intermediate, fractal, and Porod. In the fractal region, the scattering intensity is approximated well by the non-coherent sum of intensities of all disks. The higher the ratio ${L}/r_0$ is, the better the approximation (\ref{eq:incohsas}) works~\cite{chernyARXIVE16}.  Thus, the correlations between spatial positions of the disks can play a role, but they lead only to additional oscillations, while the value of the scattering exponent is preserved, as one can see from Fig.~\ref{fig:fig2}. Note that the oscillations would be smeared and not visible at all in practical experimental measurements.

\section{\label{sec:koch}Koch snowflakes: construction and scattering properties}
\begin{figure}[tb]
\centerline{\includegraphics[width=0.47\columnwidth]{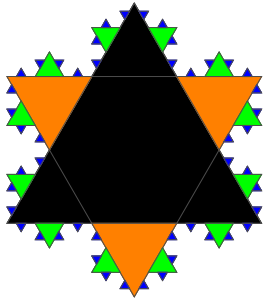}}
\caption{(Color online) Koch snowflake as a sum of various iterations of mass fractals shown in different colors. The second iteration of Koch snowflake is shown, see Fig.~\ref{fig:fig4}.
}
\label{fig:fig3}
\end{figure}

The Koch snowflake (KS) is a two-dimensional surface fractal, which can be constructed as a sum of mass fractals composed of triangles, see Fig.~\ref{fig:fig3}. We start from an initial equilateral triangle (zeroth mass fractal iteration) with edge $a$ and area $S^\mathrm{T}=\sqrt{3}a^2/4$. Then each edge is divided into three segments of equal length $a/3$, and an outward equilateral triangle is added with the base coinciding with the center segment. Then the operation is applied repeatedly to each line segment. After $m$th iterations, the total number of equal edges of length $a_m=a/3^m$ is equal to $3\cdot 4^m$. Therefore, the Hausdorff dimension of the perimeter is easily calculated as $D_{\mathrm{s}}=\lim_{m\to\infty}\log (3\cdot 4^m)/\log(a/a_m)=\log 4/\log 3\simeq 1.26$, and thus we have $D_{\mathrm{s}}>1$.

The $n$th mass fractal iteration consists of triangles of equal sizes with the edge $a_n=a/3^n$, and their number is equal to $N_n=3\cdot 4^{n-1}$ for $n=1,2,\ldots$. Note that the dimensions of the \emph{mass fractal} and the \emph{perimeter} coincide and equal $D_{\mathrm{s}}=\log 4/\log 3$. The area of the $n$th mass fractal iteration is $N_n S_\mathrm{T}/3^{2n}$.

Now it is more convenient to consider the star of David as the initiator KS (the zeroth iteration). Then the total area of KS at the $m$th iteration is given by
\begin{equation}
S_m=S^\mathrm{T}+\sum^{m+1}_{n=1}\frac{S^\mathrm{T}}{3^{2n}}N_n=\frac{4 S^\mathrm{T}}{5} \left(2 - \frac{1}{3} \frac{4^m}{9^m} \right).
\label{KS_area}
\end{equation}
The area $S^\mathrm{KS}$ of the ideal KS is obtained from this equation in the limit $m\to\infty$, which leads to $S^\mathrm{KS}=8 S^\mathrm{T}/5=2\sqrt{3}a^2/5$.

\begin{figure}[tb]
\begin{center}
\includegraphics[width=\columnwidth]{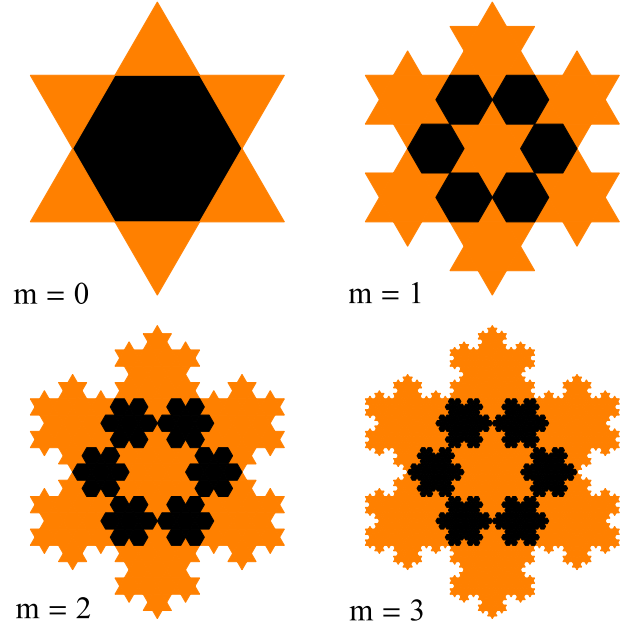}
\end{center}\caption{(Color online) A construction algorithm of the Koch snowflake. The generator ($m=0$) is built from six triangles (orange) and one hexagon (black). The first iteration is obtained with six hexagons (black) and seven zeroth iterations (orange), scaled with factor $1/3$. For constructing the second iterations, we take the first iteration ($m=1$) and subtract the six outside zeroth iterations, thus obtaining a ``modified hexagon". The second iteration ($m=2$) is composed of six  ``modified hexagons" (black) and seven first iterations (orange), scaled with factor $1/3$. The third iteration is constructed in the same manner.}
\label{fig:fig4}
\end{figure}

The standard algorithm of constructing KS, described above, is not convenient for obtaining the KS scattering amplitude, because it is not simple technically to calculate analytically the positions of the triangles for arbitrary iteration. For this reason, we adopt here a slightly modified algorithm of the paper \cite{burns94}, see Fig.~\ref{fig:fig4}. With this algorithm, the recurrence formula for the scattering amplitude $A_m(\bm{q})\equiv S_m F_m(\bm{q})$ of the $m$th iteration of KS can be write down by using the properties of the scattering amplitudes discussed in Sec.~\ref{sec:Theory}.
\begin{align}
A_m(\bm{q})\!=&6G_2(\bm{q})[\bs^2A_{m-1}(\bs\bm{q})-6 G_1(\bs\bm{q})\bs^4A_{m-2}(\bs^2\bm{q})]\nonumber\\
&+\bs^2A_{m-1}(\bs\bm{q})[1+6G_1(\bm{q})]\label{AmKS}
\end{align}
where the scaling factor takes the value $\bs=1/3$, and $G_1(\bm{q}) = \frac{1}{6}\sum_{j=0}^{5} e^{-i\bm{q}\cdot \bm{c}_{j}}$, $G_2(\bm{q}) = \frac{1}{6}\sum_{j=0}^{5} e^{-i\bm{q}\cdot \bm{b}_{j}}$ with the translation vectors $\bm{b}_{j} = \frac{2a}{3\sqrt{3}}\{\cos(\pi j/3),~\sin(\pi j/3) \}$ and $\bm{c}_{j} = \frac{2a}{9}\{ \cos\left(\pi (j+1/2)/3 \right), \sin\left( \pi(j+1/2)/3 \right) \}$.

\begin{figure}[tb]
\begin{center}
\includegraphics[width=\columnwidth]{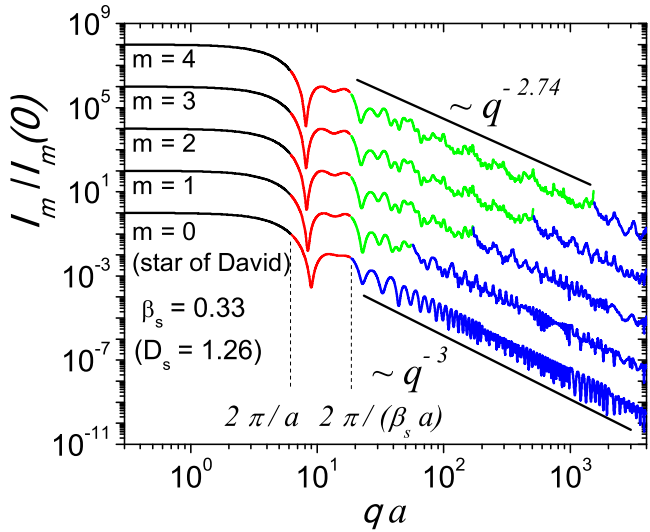}
\end{center}\caption{(Color online)  Scattering intensity for the first four iterations of the monodisperse Koch snowflake. Scattering curve for the $m$th iteration is scaled up for clarity by the factor $10^{2m}$. The Guinier, intermediate, fractal, and Porod regions are shown in black, red, green, and blue, respectively.}
\label{fig:fig5}
\end{figure}

Equation (\ref{AmKS}) allows us to obtain the nonnormalized scattering amplitude of KS for arbitrary iteration provided the amplitudes of the zeroth and first iterations are known. They can be calculated from the scattering amplitudes of triangles as discussed in Appendix~\ref{TrAmpl}. The scattering intensity is proportional to the squared amplitude averaged with respect to the polar angle: $I_m(q)\sim \langle |A_m(\bm{q})|^2\rangle$.

Equation (\ref{AmKS}) can be used to derive the recurrence relations for the area and gyration radius of KS. Indeed, taking this equation at $q=0$, we have $A_m(0)=S_m$, and $G_1(0)=G_2(0)=1$. Substituting $\bs=1/3$ yields the recurrence relation for the area $S_{m}=(13S_{m-1} - 4 S_{m-2})/9$, which is consistent well with the explicit formula (\ref{KS_area}).

The radius of gyration $R_{m}$ of the $m$th iteration of KS is determined from the expansion of scattering intensity $I(q) = I(0)(1-q^{2}R_{g}^{2}/2 + \cdots)$ for $q \rightarrow 0$ at $d=2$~\cite{svergun87:book}. Thus, we obtain $I_m(q)/I_m(0)=\langle |A_m(\bm{q})|^2\rangle/S_m^2=(1-q^{2}R_{m}^{2}/2+\cdots)$. If a structure has a rotational symmetry of order $n\geqslant3$  with respect to the center-of-mass, then the scattering amplitude is rotationally symmetric at small $\bm{q}$ up to quadratic terms. KS is invariant under rotation through the angle $\pi/3$ about the center, which implies that the rotational symmetry has the order $n=6$. This gives us $A_m(\bm{q})=S_m(1-q^{2}R_{m}^{2}/4+\cdots)$. Substituting this relation and the expansions for $G_1(q)=1-a^2q^2/81+\cdots$ and $G_2(q)=1-a^2q^2/27+\cdots$ into Eq.~(\ref{AmKS}) yields
\begin{align}
&R_m^2=\nonumber\\
&\frac{351 R_{m-1}^2S_{m-1}-12 R_{m-2}^2S_{m-2} + 32 a^2 (9 S_{m-1}-2 S_{m-2})}{2187 S_m}
\label{RgKS}
\end{align}
The radii of gyration for the first two iterations can be calculated straightforwardly: $R_0^2=11 a^2/108$ (star of David) and $R_1^2=223a^2/1944$. The radius of gyration of the ideal KS can easily be obtained from Eq.~(\ref{RgKS}). In the limit $m\to \infty$, the area $S_m$ tends to the area $S^\mathrm{KS}$ of the ideal KS, and $R_m\to R_\mathrm{KS}$. Then by taking the limit from the both sides of Eq.~(\ref{RgKS}) and cancelling $S^\mathrm{KS}$ from the numerator and denominator, we arrive at the linear equation for $R^2_\mathrm{KS}$, which yields $R^2_\mathrm{KS}=4 a^2/33$, and, hence,
\begin{align}
R_\mathrm{KS}=2a/\sqrt{33}.
\label{RgKSideal}
\end{align}
The scattering intensities are shown in Fig.~\ref{fig:fig5}. Generally, all the properties of the scattering curves are the same as in the case of Cantor-like surface fractals, presented above in Sec.~\ref{subsec:monoformfactor}. Since the overall size of KS is of order of $a$, the upper border of the Guinier range is about $2\pi/a$. The Fractal range lies between $2\pi/(a\bs)$ and $2\pi/(a\bs^{m+1})$, because the edge of smallest triangles equals $a\bs^{m+1}$.

Let us consider the contribution of different mass fractal amplitudes to the total scattering intensity of KS. Because of additivity of the scattering amplitude, each mass fractal amplitude can be calculated as the difference between amplitudes of two consecutive iterations of KS (see Fig.~\ref{fig:fig3})
\begin{equation}\label{mass_ampl}
M_m(\bm{q})=A_{m-1}(\bm{q}) - A_{m-2}(\bm{q}),
\end{equation}
where $m=2,3,\ldots$. The zeroth mass fractal iteration is the largest triangle in Fig.~\ref{fig:fig3}, so we have $M_0(\bm{q}) = A^{\mathrm{T}}(\bm{q}) = S^{\mathrm{T}} F^{\mathrm{T}}(\bm{q})$, and the first mass fractal amplitude is given by $M_1(\bm{q}) = A_{0}(\bm{q})-A^{\mathrm{T}}(\bm{q})$.

\begin{figure}[tb]
\centerline{\includegraphics[width=\columnwidth]{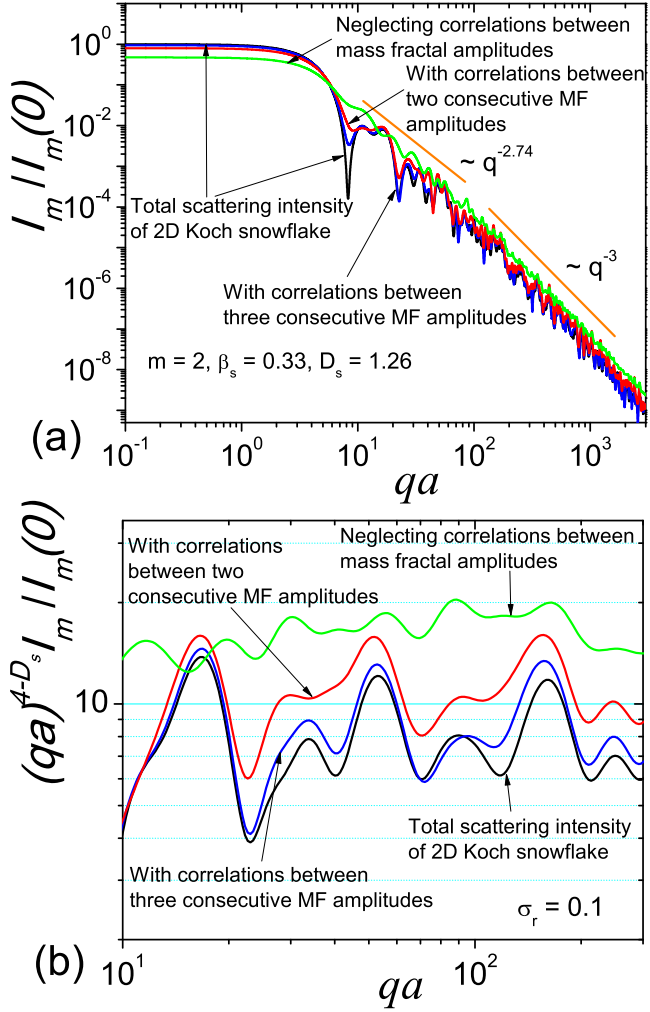}}
\caption{(Color online) SAS from KS and its approximations by the scattering from the mass fractal composing KS (see Fig.~\ref{fig:fig3}). (a) The total intensity (black) and the intensities taking into account various correlations between the mass fractal amplitudes. The correlations between \emph{three} consecutive mass fractal amplitudes are included (blue), and the same for the correlations between \emph{two} consecutive mass fractal amplitudes (red). Neglecting all the correlations between mass fractals (green) is not good enough for describing the total intensity of KS. (b) Approximate log-periodicity of the curve $I(q)q^{4-D_{\mathrm{s}}}$  with the period $\bs=1/3$. The polydisperse scattering intensities are shown for the relative variance $\sigma_{\mathrm{r}} = 0.1$. One can observe an interference between different mass fractal amplitudes, so their correlations are important.}
\label{fig:fig6}
\end{figure}

Inversely, one can write the KS amplitude as a sum of the mass fractal amplitudes
\begin{equation}\label{KS_ampl_mass}
A_m(\bm{q})=\sum_{n=0}^{m+1} M_{n}(\bm{q}).
\end{equation}
Then the KS intensity $I_m(q)=\langle|A_m(\bm{q})|^2\rangle$ contains not only the mass fractal intensities $\langle|M_n(\bm{q})|^2\rangle$ but the correlations between the mass fractal amplitudes
\begin{align}
I_m&(q)=\sum_{n=0}^{m+1} \langle|M_{n}(\bm{q})|^2\rangle \nonumber\\
&+\sum_{0\leqslant n<p\leqslant m+1} \langle M^*_{n}(\bm{q})M_{p}(\bm{q})+M_{n}(\bm{q})M^*_{p}(\bm{q})\rangle.
\label{int_ampl_mass}
\end{align}
One can neglect the non-diagonal (interference) terms in this equation and even more, completely neglect the interference between the amplitudes of triangles composing the mass fractals. This approximations work well frequently, say, for the Cantor surface fractal (see Sec.~\ref{subsec:monoformfactor} above). However, this scheme does not work properly for the KS, see Fig.~\ref{fig:fig6}. The reason is that distances between  different mass fractal iterations and between triangles within one mass fractal iteration can be of order of their sizes, and we have to take into account the interference  terms in Eq.~(\ref{int_ampl_mass}).

Nevertheless, one can reduce the problem, in effect, to the incoherent sum of the ``combined" mass fractals.
Indeed, considering the correlations between two consecutive mass fractal iterations like $\langle M^*_{0}M_{1}\rangle$, $\langle M^*_{1}M_{2}\rangle$, and so on, and neglecting the other correlations, we obtain from Eq.~(\ref{int_ampl_mass})
\begin{align}
I_m(q)\simeq\sum_{n=0}^{m} \langle|M_{n}(\bm{q})+M_{n+1}(\bm{q})|^2\rangle -\sum_{n=1}^{m} \langle|M_{n}(\bm{q})|^2\rangle
\label{int_double_mass}
\end{align}
SAS from the surface Cantor fractal is described well by incoherent sum of \emph{single} mass fractal intensities, while the first sum in the approximation (\ref{int_double_mass}) is nothing else but \emph{incoherent sum} of intensities of \emph{pairs} of consecutive amplitudes. The SAS intensities of each pair behave like a mass fractal with the power-law decay $I(q)\sim q^{-D_\mathrm{m}}$ at $D_\mathrm{m}=\ds$, which results in the power-law decay of the intensity (\ref{int_double_mass}) $I(q)\sim q^{D_\mathrm{s}-2d}$ with $d=2$ for the plane.

By analogy with the pair consecutive amplitudes, one can further improve the approximation (\ref{int_double_mass}) for the SAS intensity by including the triple consecutive amplitudes $\langle|M_{n}+M_{n+1}+M_{n+2}|^2\rangle$. The results for the KS is shown in Fig.~\ref{fig:fig6}a.

One of the main properties is the approximate log-periodicity of the curve $I(q)q^{4-D_{\mathrm{s}}}$ within the fractal range, which is illustrated in Fig.~\ref{fig:fig6}b. As one can see, complete ignorance of correlations between the mass fractal amplitudes leads to a bad approximation.

\section{\label{sec:Conclusions}Conclusions}
The construction algorithm for the Koch snowflake (see Fig.~\ref{fig:fig4}) allows us to write down the recurrence relation (\ref{AmKS}) for the scattering amplitude. The analytical expression for the scattering amplitude of $2D$ Cantor-like surface fractal is derived. We obtain the recurrence formula for the radius of gyration of the Koch snowflake (\ref{RgKS}), which yields the radius of gyration (\ref{RgKSideal}) of the ideal Koch snowflake.

It is shown that at a given iteration $m$, the both surface fractal models can be represented as a sum of mass fractals at iterations from zero to $m$. This confirms that in general, \emph{any surface fractal can be represented as a sum of mass fractals}. While the ``rough structure" of SAS (including the borders of fractal region and the power exponent $\ds-2d$) is determined by the power-law distribution of the triangle sizes, the superimposed interference structure of the intensity needs more precise approximations.

It is shown that for the Cantor-like surface fractal, the correlations between mass fractal amplitudes can be neglected, however for the Koch snowflake the correlations between amplitudes are important. The reason is that distances between  different mass fractal iterations and between triangles within one mass fractal iteration can be of order of their sizes, and we have to take into account some of the interference terms in Eq.~(\ref{int_ampl_mass}). Then the most important interference terms can be built in the model with Eq.~(\ref{int_double_mass}). The log-periodicity of he curve $I(q)q^{4-D_{\mathrm{s}}}$, where $1 < D_{\mathrm{s}} < 2$, arises from the self-similarity of \emph{sizes} of basic structural units, in contrast with mass fractals, where the log-periodicity arises from the self-similarity of \emph{distances} between structural units.

The present analysis might be useful for obtaining structural information (overall dimension of the fractal, size of the smallest structural unit composing the fractal, the fractal iteration number, and the scaling factor) from various artificially prepared nano and micro systems, such as for the recently obtained molecular Sierpinski hexagonal gasket incorporating the Star of David and the Koch snowflake motifs~\cite{newkome06} or the three-dimensional analog of the Koch snowflake~\cite{berenschot13}.

\acknowledgments
The authors acknowledge financial support from JINR--IFIN-HH projects. A.I.K. acknowledges Russian program ``5Top100" of the Ministry of Education and Science of the Russian Federation.

\appendix
\section{Scattering from a triangle, hexagon and DS}
\label{TrAmpl}

Consider an isosceles triangle with the altitude $h$ and the length of its base $a$. The area of triangle is equal to $S^{\mathrm{T}}=ah/2$. We choose the Cartesian coordinate system where the base is parallel to the $x$-axes and the opposite vertex coincides with the origin. Then the normalized scattering amplitude is obtained with Eq.~\eqref{eq:normff}, which becomes now a surface integral
\begin{equation}
F^{\mathrm{T}}(\bm{q})=\frac{1}{S^{\mathrm{T}}}\int_{0}^{a}\d y\int_{-\frac{ya}{2h}}^{\frac{ya}{2h}}\d x\,e^{-i( x q_{x}+y q_{y} )}
\label{eq:normfftriangle}
\end{equation}
with the scattering vector $\bm{q}=\{q_{x},q_{y}\}$. By calculating the integral, we arrive at the analytical expression
\begin{equation}
F^{\mathrm{T}}(\bm{q}) = \frac{2e^{-i\alpha} \left( \beta e^{i\alpha}-\beta\cos\beta-i\alpha\sin\beta \right)}{\beta\left( \beta^{2}-\alpha^{2} \right)},
\label{eq:normfftrianglefinal}
\end{equation}
where we put by definition $\alpha\equiv hq_{y}$, $\beta\equiv aq_{x}/2$. For an equilateral triangle, we have $h= a\sqrt{3}/2$.

The scattering amplitude of any geometrical set, composed of triangles, can be obtained by summing the triangle amplitudes, which are appropriately scaled, rotated, and translated. Then a hexagon can be constructed from the six equilateral triangles, and a star of David can be composed of one big equilateral triangle and three similar triangles scaled with the factor one third, see Fig.~\ref{fig:fig3}. The formulas for their scattering amplitudes are obvious, and we do not write them down explicitly.

Note that hexagon and star of David have the inversion symmetry $\bm{r}\to -\bm{r}$ with respect to their center-of-masses, and, thus, their amplitudes are real $A^*(\bm{q})=A(\bm{q})$, provided the coordinate origin is chosen in the centers.

\bibliography{sas_sf1}

\end{document}